# Effects of Slotted Structures on Nonlinear Characteristics of Natural Convection in a Cylinder with an Internal Concentric Slotted Annulus

Chunyun Shen[a], Mo Yang[a1], Yuwen Zhang[b], Zheng Li[a,b]

a. School of Energy and Power Engineering, University of Shanghai for Science and Technology, Shanghai 200093, China
b. Department of Mechanical and Aerospace Engineering, University of Missouri, Columbia, MO 65211, USA

## Abstract

Natural convection in a cylinder with an internally slotted annulus was solved by SIMPLE algorithm, and the effects of different slotted structures on nonlinear characteristics of natural convection were investigated. The results show that the equivalent thermal conductivity $K_{eq}$ increases with Rayleigh number, and reaches the maximum in the vertical orientation. Nonlinear results were obtained by simulating the fluid flow at different conditions. With increasing Rayleigh number, heat transfer is intensified and the state of heat transfer changes from the steady to unsteady. We investigated different slotted structures effects on natural convection, and analyze the corresponding nonlinear characteristics.

## NOMENCLATURE

$c_p$   specific heat (J/kgK)

$F$   non-dimensional time

$g$   gravitational acceleration (m/s$^2$)

$p$   pressure (Pa)

$P$   non-dimensional pressure
$Pr$   Prandtl number

$Q$   heat generation by the heaters (W/m)

$Ra$   Rayleigh number

---

[1] Corresponding author. Email: yangm@usst.edu.cn. Tel:+86 13024171865

$r$ radial coordinate

$R$ non-dimensional radial coordinate

$r_i$ radius of inner slotted cylinder (m)

$r_o$ radius of outer cylinder (m)

$S$ non-dimensional internal heat generation

$t$ time (s)

$T$ temperature (K)

$T_o$ temperature of outer walls (K)

$T_i$ temperature of inner walls (K)

$U$ non-dimensional velocity in the radial coordinate

$v_r$ radial velocity

$v_\theta$ tangential velocity

$V$ non-dimensional velocity in the tangential coordinate

Greek symbols

$\phi$ angle of slot from the vertical (rad)

$\mu$ viscosity (Kg/ms)

$\rho$ density (Kg/m$^3$)

$\Theta$ non-dimensional temperature

$\theta$ angular coordinate

$\Delta\tau$ non-dimensional time step

$\sigma$ thermal diffusivity (m$^2$/s)

$\beta$ thermal expansion coefficient (1/K)

$\delta$ thickness of inner cylinder

# 1. Introduction

Nonlinear dynamics of natural convection in the closed cavity have attracted increasing interests in the recent years. Quere et al. [1-3] studied natural convection nonlinear phenomena in the square cavity using a spectral method. Li et al. [4-6] studied natural convection in the square cavity using various numerical methods. Liu and Tao [7] investigated periodic oscillations in a closed rectangular cavity, and their results showed that at $Ra=2.0\times10^4$ the system was steady symmetric, at $Ra=2.0\times10^5$ to $10^6$ the system performed non-symmetrical oscillation, and at Ra=2.0×10$^6$, the flow turned into counter-periodical oscillation. Liaqat and Baytas [8] employed SIMPLER algorithm to simulate the natural convection in a square cavity with internal heat source at high Rayleigh number. Their results reached periodic oscillation for the case of $Ra<10^7$ and non-periodic oscillation when $Ra>10^8$. Fusegi et al. [9] used SIMPLE algorithm with QUICK scheme to simulate natural convection in a three-dimensional square cavity. Li et al [10] discussed the nonlinear results of natural convection melting problems in a square cavity. Ishida et al. [11] and Benouaguef et al. [12] studied nonlinear characteristics of natural convection in inflatable square cavity using different numerical methods.

Bishop [13] and Powe et al. [14] experimentally studied the natural convection in a closed helium filled annular space. Yoo [15] reported that when Rayleigh number was greater than a critical value, the stable solution was observed. Rao et al. [16] investigated heat convection in the annular space, and obtained the critical Rayleigh number for the flow regime transition under different conditions. Net et al. [17], Mizushima et al. [18] studied the development processes of flow and the change of heat transfer with time, and discussed the transition from steady-state to chaos in detail. Fant et al. [19-20] studied the flow and heat transfer in a two-dimensional annular space. Yoo [21-22] studied the nonlinear phenomena occurred in a narrow annulus at different parameters.

Kuleek [23] studied heat transfer in a circle with an internal slotted annulus and proposed an approximate formula. Wang et al. [24] conducted experiments to investigate natural convection in the current bus. Zhang et al. [25-26] experimentally studied the horizontal cylinder with slotted octagonal structure. Yang et al. [27] discussed multiple solutions for natural convection in the system. In this paper, we will investigate the slotted structure effects on the flow and heat transfer, and analyze its influence on the nonlinear characteristics.

## 2. Physical model

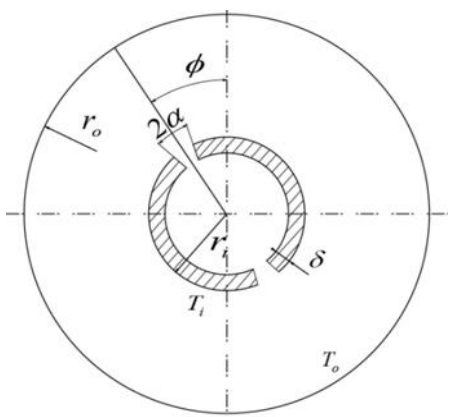

Fig. 1 The closed circle within slotted round mathematical model

Figure 1 shows the physical model of the problem under consideration. It is assumed that the natural convection is two-dimensional, and the inner and outer surfaces are kept at constant temperatures of $T_i$ and $T_o$ ($T_i > T_o$). And $\phi$ is the slotted direction deflection angle, $\phi=90°$ is for the horizontal slot. The other geometric parameters are: $\eta = r_i / r_o = 5/13$, $S = 2\alpha / \pi = 0.1$, and $\delta = 0.1125$. Working fluid is incompressible, and Boussinesq assumption is employed. Non-slip boundary conditions are used for all solid walls, and viscous dissipation effect in the energy equation is neglected.

This problem can be described by the following equations:

Continuity equation:

$$\frac{1}{r}\frac{\partial}{\partial r}(r \cdot v_r) + \frac{1}{r}\frac{\partial v_\theta}{\partial \theta} = 0 \qquad (1)$$

Momentum equations:

$$\frac{\partial v_\theta}{\partial t} + v_r \frac{\partial v_\theta}{\partial r} + \frac{v_\theta}{r}\frac{\partial v_\theta}{\partial \theta} + \frac{v_\theta \cdot v_r}{r} = -\frac{1}{\rho r}\frac{\partial p'}{\partial \theta} - \beta \cdot g \cdot (T-T_r)\cdot \sin\theta + \nu \cdot \left[\frac{\partial^2 v_\theta}{\partial r^2} + \frac{1}{r}\frac{\partial v_\theta}{\partial r} + \frac{1}{r^2}\frac{\partial^2 v_\theta}{\partial \theta^2} - \frac{v_\theta}{r^2} + \frac{2}{r^2}\frac{\partial v_r}{\partial \theta}\right] \qquad (2)$$

$$\frac{\partial v_r}{\partial t} + v_r \frac{\partial v_r}{\partial r} + \frac{v_\theta}{r}\frac{\partial v_r}{\partial \theta} - \frac{v_\theta^2}{r} = -\frac{1}{\rho}\frac{\partial p'}{\partial r} + \beta \cdot g \cdot (T-T_r)\cdot \cos\theta + \nu \cdot \left[\frac{\partial^2 v_r}{\partial r^2} + \frac{1}{r}\frac{\partial v_r}{\partial r} + \frac{1}{r^2}\frac{\partial^2 v_r}{\partial \theta^2} - \frac{v_r}{r^2} - \frac{2}{r^2}\frac{\partial v_\theta}{\partial \theta}\right] \qquad (3)$$

Energy equation:

$$\frac{\partial T}{\partial t} + v_r \frac{\partial T}{\partial r} + \frac{v_\theta}{r}\frac{\partial T}{\partial \theta} = a \cdot \left(\frac{\partial^2 T}{\partial r^2} + \frac{1}{r}\frac{\partial T}{\partial r} + \frac{1}{r^2}\frac{\partial^2 T}{\partial \theta^2}\right) \qquad (4)$$

Equations (1) – (4) are subject to the following boundary and initial conditions:

$r_i - \delta \leq r \leq r_i, \alpha \leq \theta \leq \pi - \alpha$ or

$$\pi + \alpha \leq \theta \leq 2\pi - \alpha : v_r = v_\theta = 0, \quad T = T_h \qquad (5)$$

$$r = r_o : v_r = v_\theta = 0, \quad T = T_c \qquad (6)$$

Defining the following non-dimensional variables:

$R = r/L$, $F = at/L^2$, $U = v_\theta/u_R$, $V = v_r/u_R$, $P^* = P/(\rho u_R^2)$,

$$\Theta = (T-T_c)/(T_h - T_c), \quad Ra = \beta g L^3 (T_h - T_c)/(a \cdot \gamma), \quad U_R = (Ra \cdot \Pr)^{(1/2)} (a/L) \qquad (7)$$

Equations (1) - (4) become:

$$\frac{\partial V}{\partial R} + \frac{V}{R} + \frac{1}{R}\frac{\partial U}{\partial \theta} = 0 \qquad (8)$$

$$\frac{\partial V}{\partial F} + V\frac{\partial V}{\partial R} + \frac{U}{R}\frac{\partial V}{\partial \theta} = -\frac{\partial P}{\partial R} + \frac{\Pr}{(Ra\,\Pr)^{1/2}}\left[\frac{1}{R}\frac{\partial}{\partial R}\left(R\frac{\partial V}{\partial R}\right) + \frac{1}{R^2}\frac{\partial^2 V}{\partial \theta^2}\right] + S_R \qquad (9)$$

$$\frac{\partial U}{\partial F} + V\frac{\partial U}{\partial R} + \frac{U}{R}\frac{\partial U}{\partial \theta} = -\frac{1}{R}\frac{\partial P}{\partial \theta} + \frac{\Pr}{(Ra\,\Pr)^{1/2}}\left[\frac{1}{R}\frac{\partial}{\partial R}\left(R\frac{\partial U}{\partial R}\right) + \frac{1}{R^2}\frac{\partial^2 U}{\partial \theta^2}\right] + S_\theta \qquad (10)$$

$$\frac{\partial \Theta}{\partial F} + V\frac{\partial \Theta}{\partial R} + \frac{U}{R}\frac{\partial \Theta}{\partial \theta} = \frac{1}{(Ra\,\Pr)^{1/2}}\left[\frac{1}{R}\frac{\partial}{\partial R}\left(R\frac{\partial \Theta}{\partial R}\right) + \frac{1}{R^2}\frac{\partial^2 \Theta}{\partial \theta^2}\right] \qquad (11)$$

where:

$$S_\theta = -\frac{UV}{R} + \frac{\Pr}{(Ra\,\Pr)^{1/2}}\left(-\frac{U}{R^2} + \frac{2}{R^2}\frac{\partial V}{\partial \theta}\right) - \Theta \sin\theta \qquad (12)$$

$$S_R = -\frac{U^2}{R} + \frac{\Pr}{(Ra\,\Pr)^{1/2}}\left(-\frac{V}{R^2} + \frac{2}{R^2}\frac{\partial U}{\partial \theta}\right) + \Theta \cos\theta \qquad (13)$$

The corresponding periodic boundary conditions are:

$$U(\theta=0,R)=U(\theta=2\pi,R), \quad V(\theta=0,R)=V(\theta=2\pi,R),$$

$$\Theta(\theta=0,R)=\Theta(\theta=2\pi,R) \tag{14}$$

For the inner slotted annulus, the following conditions must be satisfied:

$$R_i - \delta/L \leq R \leq R_i, \alpha \leq \theta \leq \pi - \alpha, U = V = 0, \Theta = 1$$

$$R_i - \delta/L \leq R \leq R_i, \gamma + \pi \leq \theta \leq 2\pi - \alpha, U = V = 0, \Theta = 1 \tag{15}$$

The initial conditions are:

$$F = 0, \ U = V = \Theta = 0 \tag{16}$$

In order to describe the strength of the heat transfer under different Rayleigh number, the average equivalent thermal conductivity is defined as:

$$K_{eq} = \frac{Q}{2\pi(T_i - T_o)} \ln \frac{r_o}{r_i} \tag{17}$$

The whole heat transfer area includes the entire inner surface of outer cylinder and two half-annular surfaces. To compare with the existing experimental results [21], the revised average thermal conductivity can be defined as:

$$K_{eqs} = \frac{K_{eq}}{1-S}, \quad S = 2\alpha/\pi \tag{18}$$

## 3. Results and discussions

## 3.1 Numerical method assessment

**Table 1 The assessment of the grid quality**

| grid number / Ra | 60×40 | 80×40 | 100×50 | 150×50 | 190×90 |
|---|---|---|---|---|---|
| 1×10$^5$ | 4.862 | 4.840 | 4.801 | 4.798 | 4.796 |
| 1×10$^6$ | 9.107 | 9.006 | 8.945 | 8.939 | 8.936 |

We can obtain the best grid number and the more accurate results through grid-independent test. At the time step of Δt = 0.01, the $K_{eq}$ obtained with different grid numbers and the results are shown in Table 1. It can be seen from Table 1 that with increasing grid number $K_{eq}$ decreases. But when the grid number reach to 100(θ)×50(r), $K_{eq}$ changes slowly and therefore, we consider that the results reached grid number-independent; thus, the grid number used in the rest of simulation will be 100(θ)×50(r). The time step-independent test is then carried out with the grid number of $100(\theta) \times 50(r)$. Table 2 shows the relationship between time step and $K_{eq}$ under different $Ra$. It can be seen that $K_{eq}$ decreases with decreasing time step; when $\Delta t$ reaches to 0.01, $K_{eq}$ changes insignificantly with the time step. Therefore, this time step will be used in this work.

**Table 2 The assessment result of time step**

| $\Delta t$ (time step) \ $Ra$ | 0.001 | 0.005 | 0.008 | 0.01 | 0.05 | 0.1 |
|---|---|---|---|---|---|---|
| $1 \times 10^5$ | 4.794 | 4.796 | 4.801 | 4.811 | 4.816 | 4.823 |
| $1 \times 10^6$ | 8.889 | 8.93 | 8.939 | 8.945 | 8.948 | 0.954 |

The results will also be compared with the experiment results under the same conditions. To simulate the experimental results [24] using SIMPLE method, the geometry settings are: $r_o = 130 mm$, $r_i = 49 mm$, $\delta = 4.5 mm$, $S = 0.1$, and $r_o - r_i = 40.5$. The corresponding dimensionless parameters are: $R_o = 1.625$, $R_i = 0.625$, and $\delta = 0.1125$. The empirical correlation of $K_{eqs}$ is [24]:

$$Keqs = \left[1 + (0.0894 Ra^{0.3478})^{7.884}\right]^{\frac{1}{7.884}} \quad 10^2 < Ra \leq 4.5 \times 10^4 \qquad (19)$$

$$Keqs = 0.181 Ra^{0.281}, \quad 4.5 \times 10^4 < Ra \leq 10^6 \ (20) \qquad (20)$$

Figure 2 shows the comparison between simulation and experimental results. It can be seen that the simulation results and experimental results generally agreed well. Especially, when Rayleigh number is less than $6 \times 10^5$, the results from simulation and experiments are consistent. And Rayleigh number is higher than $10^6$, the deviation is less than 6%. Therefore, the natural convection characteristics can be obtained through the simulation results.

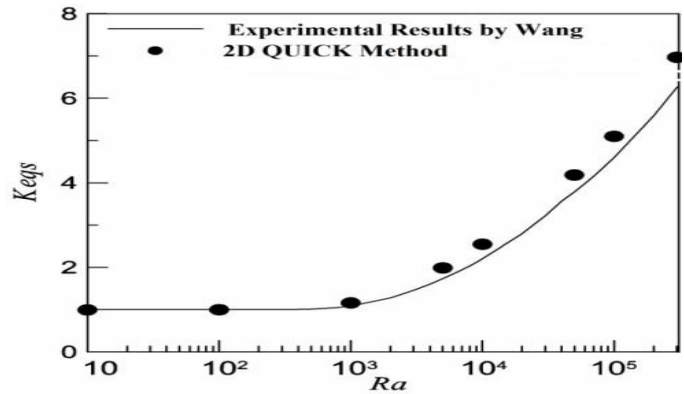

Fig. 2 Comparison with the experimental results

## 3.2 Effects of slot directions

The effective thermal conductivities at different slotted directions are calculated, and the results are shown in Figure 3. It can be seen that $K_{eqs}$ increases with $Ra$ under the same slot direction. At higher Rayleigh number, especially above $6 \times 10^5$, $K_{eqs}$ cannot reach to a stable solution and the flow showed nonlinear oscillation. Figure 4 shows $K_{eq}$ changing with time at different $Ra$, reflecting the state of heat transfer in the system under different $Ra$. At $Ra = 10^5$, $K_{eq}$ increases with time. When dimensionless time F is 160.0, $K_{eq}$ is 3.46. When *Ra* increases to $10^6$, $K_{eq}$ increases significantly before F = 150. When F = 300, $K_{eq}$ becames stable periodic fluctuations. With the increasing Rayleigh number, the periodic fluctuations disappear and begin to show non-periodic oscillation. Finally, when Rayleigh number reaches to above $5 \times 10^6$, the flow in the system become completely irregular and chaotic.

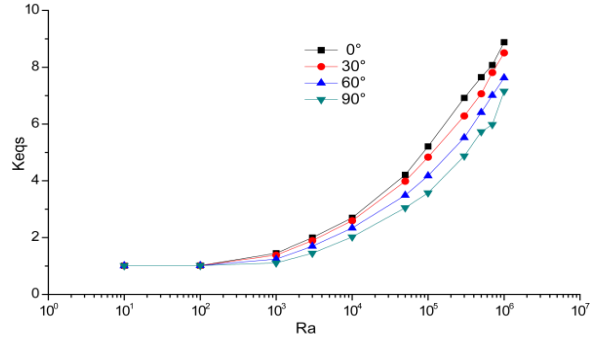

Fig. 3 $K_{eqs}$ changes with Ra

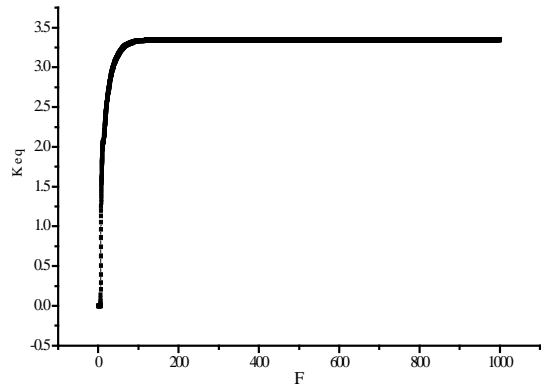

(a) $Ra = 10^5$

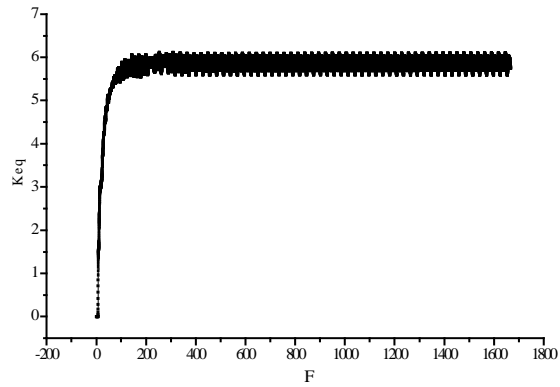

(b) $Ra = 10^6$

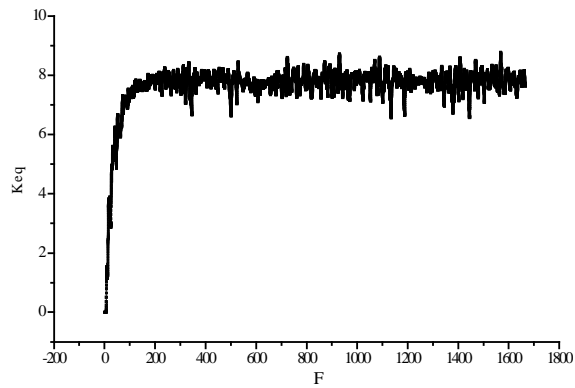

(c) $Ra = 2 \times 10^6$

Fig. 4 Variation of $K_{eq}$ in different Ra

Under the same parameters, numerical solutions can reach two or more different results. As shown in Figure 5, when Ra is $10^6$, there are three different temperature fields in the annular space. This is the static bifurcation phenomenon, which is an important characterization of the nonlinear characteristics; these results are longitudinal axisymmetric.

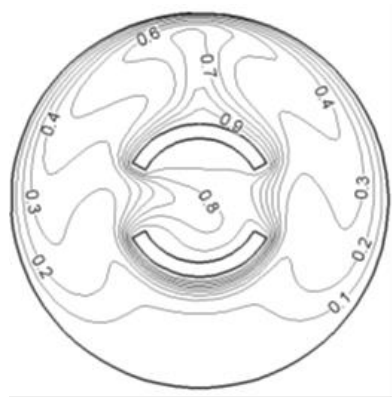

(a)

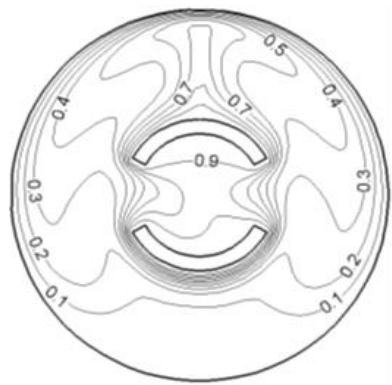

(b)

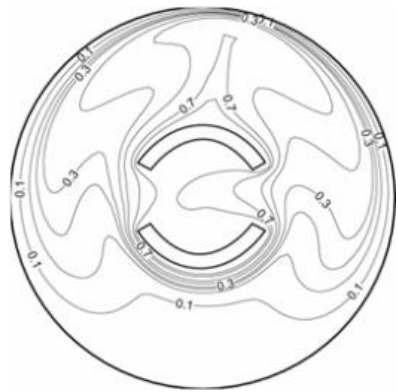

(c)

Fig. 5 Static bifurcation phenomenon

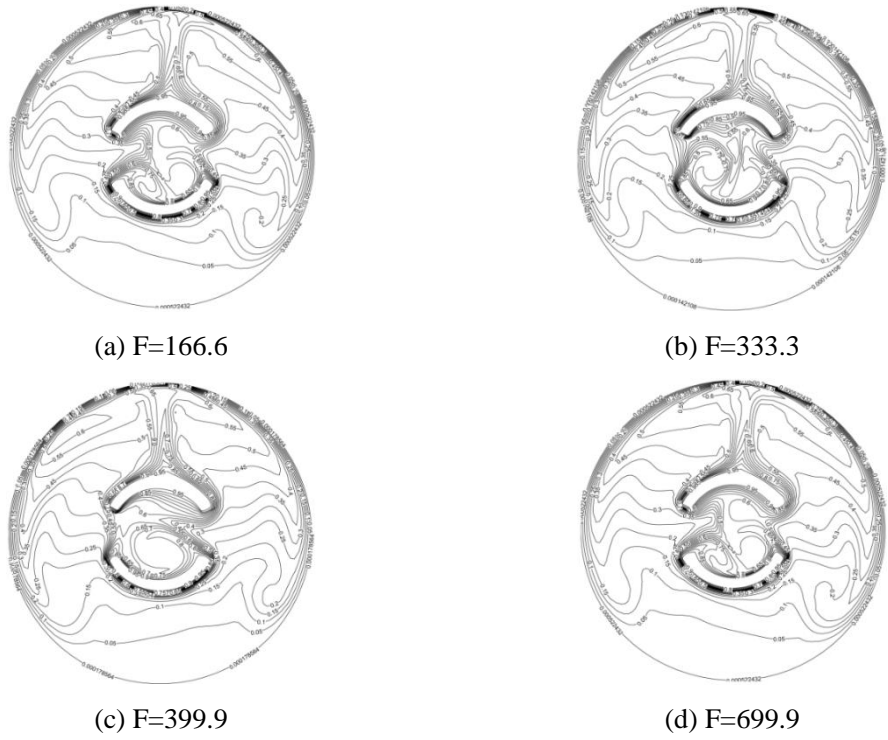

(a) F=166.6  (b) F=333.3

(c) F=399.9  (d) F=699.9

Fig. 6 Periodic oscillation phenomenon

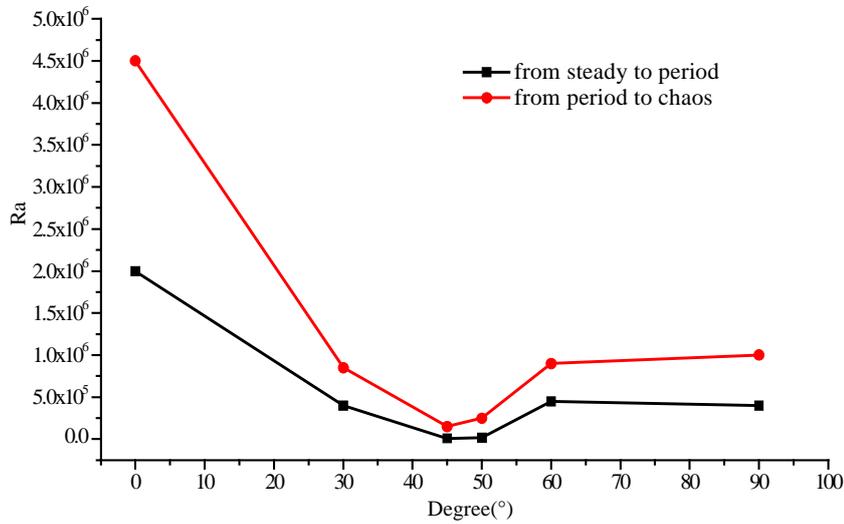

Fig. 7 Critical Ra for different slotted directions

With the increasing Rayleigh number, the state of heat transfer changes constantly. Figure 6 shows a temperature field evolution process for the case that $Ra = 10^6$. After a complete period, the temperature field returns to the initial state. In other words, the temperature field is the same for F = 699.9 and F = 166.6. Using the numerical approximation method, we obtained the critical Rayleigh numbers corresponding to the orientation of the annulus, and the results are summarized in Fig.7. It can be seen that the critical Rayleigh number are maximum, both for steady-state to period state and period state to aperiodic state, for vertical orientation; indicating heat transfer reaches the maximum, which is consistent with results of the Figure 3. The critical Rayleigh number are minimum at $\phi=45°$, showing that at this time the thermal state is easiest to change.

## 3.3 Effects of slotted degrees

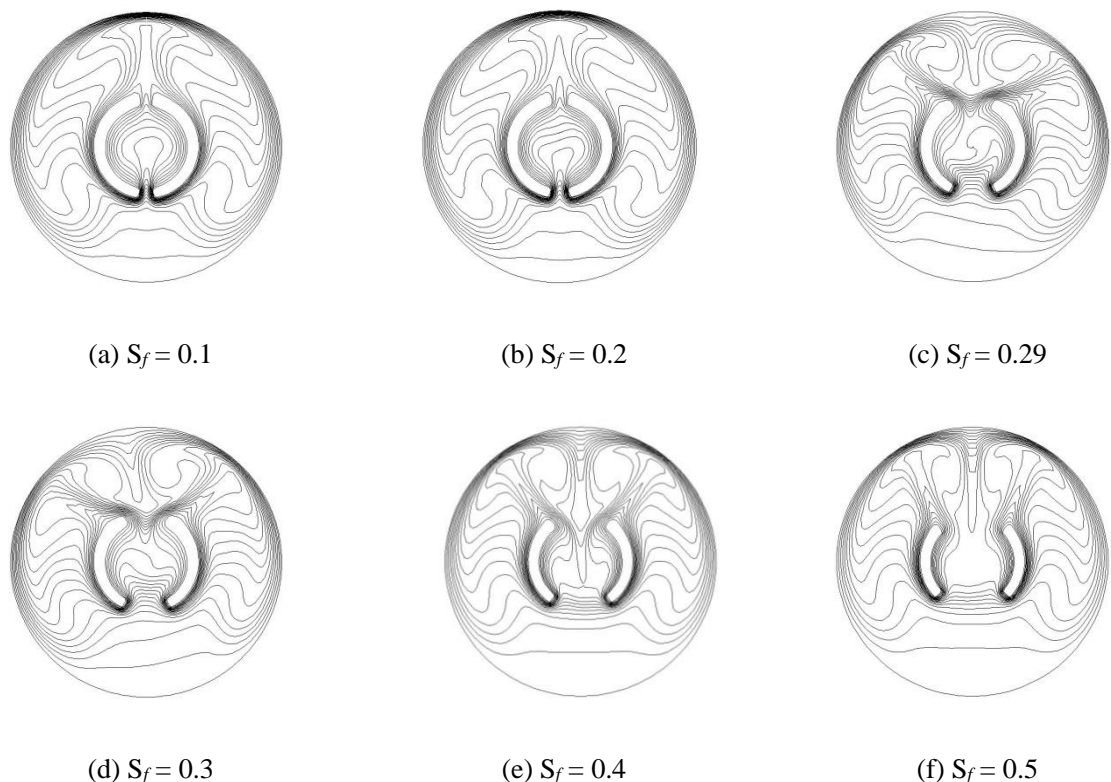

(a) $S_f = 0.1$  (b) $S_f = 0.2$  (c) $S_f = 0.29$

(d) $S_f = 0.3$  (e) $S_f = 0.4$  (f) $S_f = 0.5$

Fig. 8 Temperature fields in different slotted degrees

Figure 8 shows isotherms at different slotted degrees. The primary vortex of the temperature field is changing with increasing slotted degree. But when the slotted degree exceeds a specific value, it appears a pair of counter-clockwise vortices at the top of annular. The emergence of the vortex improves the heat convection in the circle, so the slotted cylinder can enhance the heat transfer in the system. However, the improvement is not a monotonic function of the slotted degree. Because the greater slotted degree would lead more cold fluid enter the cylinder, which prevents the heat dissipation, the heat transfer will decrease as well. Under certain radius ratio and Prandtl number, there is an optimal slotted degree at which $K_{eq}$ reaches maximum and heat transfer is also the strongest.

Figure 9 shows the change of $K_{eq}$ with critical $S_f$. It can be seen that when the annular space is at $S_f = 0.13$, natural convection is the strongest. When $Ra$ is $5 \times 10^5$, the $K_{eq}$ without slot ($S_f=0$) is 5.31. If $S_f$ is 0.13, $K_{eq}$ reaches 7.62. In other words, the heat transfer is increased by 43%.

Figure 10 shows the natural convection development process for $S_f =0.1$. The dash line and the dot line respectively represent the $K_{eq}$ development curves of the left and right half of the annulus. It can be seen from Fig. 10(a) that the $K_{eq}$ no longer changes after certain time, and eventually heat transfer reaches steady state for the case that $Ra$ is $10^6$. On the contrary, $K_{eq}$ changes to periodic oscillation as shown in Fig. 10(b) when $Ra$ is $2 \times 10^6$. Figure 11 shows the critical Rayleigh number at which the convection changing from period to aperiodicity for different slotted degrees. Due to the nonlinearity of the system, we obtained just one of the oscillation solutions. When the state of natural convection changing, the critical Rayleigh number in different slotted degrees is random, and the solutions of the system showed the nonlinear features.

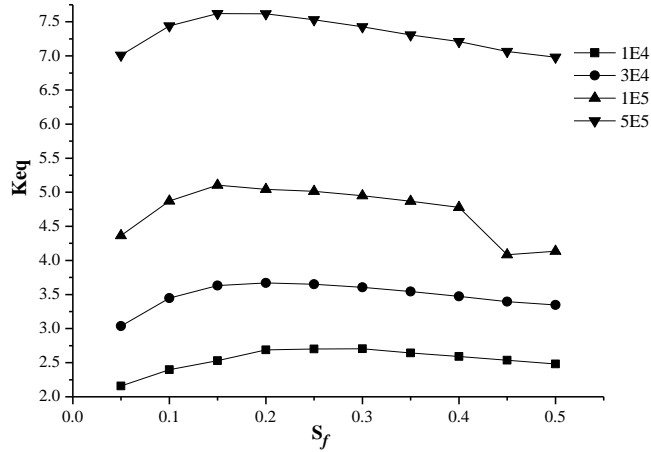

Fig. 9 $K_{eq}$ in critical $S_f$

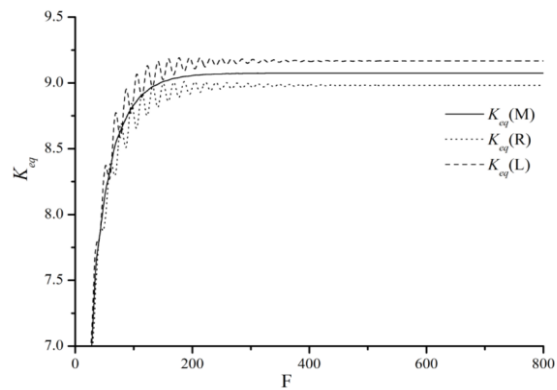

(a) $Ra = 10^6$

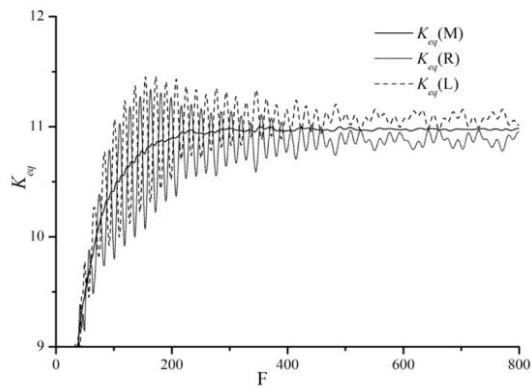

(b) $Ra = 2 \times 10^6$

Fig. 10 $K_{eq}$ changes with the dimensionless time (F)

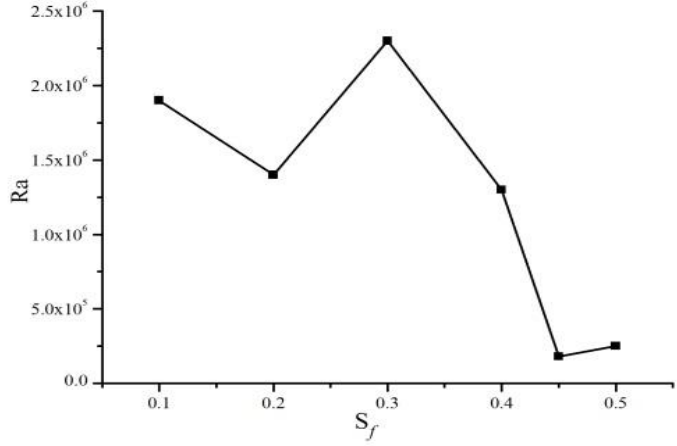

Fig. 11 Critical Ra with slotted degrees

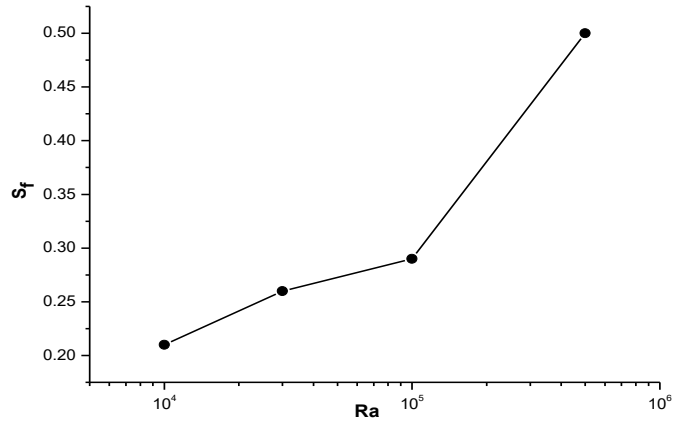

Fig. 12 Critical Ra in different $S_f$

Meanwhile, as Rayleigh number continues to increase, the temperature field became asymmetric in the annular space. Under unsteady oscillation, the semicircle vortex begins to appear alternately. Through the numerical approximation method, we obtained the critical Rayleigh number that the "secondary vortex" appears in different slotted degrees, as shown in Fig.12. When the "secondary vortex" occurred, the greater of the slotted degree, the larger of the critical Rayleigh number. In other words, appearance of the secondary vortex delays as the slotted degree increases.

## 4. Conclusions

In this paper, natural convection in a cylinder with an internally slotted annulus is simulated, and the related nonlinear problems are analyzed. The effective thermal conductivity $K_{eqs}$ increases with increasing Rayleigh number, and reaches maximum in the vertical orientation. Under certain geometric parameters, natural convection in slotted circular annulus exhibits the bifurcation phenomenon. With increasing Rayleigh number, natural convection continues to strengthen, the heat transfer state transforms from the steady to unsteady state, and finally to chaotic. The slotted inner cylinder can enhance heat dissipation, but the greater of slotted degree does not meaning the better heat transfer. There is an optimal slotted degree that $K_{eq}$ reaches to the maximum, and the system heat transfer is the strongest.

# Acknowledgments

This work is supported by Chinese National Natural Science Funds (Grant number 51476103), Innovation Program of Shanghai Municipal Education Commission (Grant number 14ZZ134)，The Innovation Fund Project for Graduate Student of Shanghai (Grant number JWCXSL1401).